\documentclass[fleqn]{article}
\usepackage{amsmath}
\usepackage{amsfonts,amssymb,cite}
\usepackage{epsf,graphicx}
\def\noi{\noindent}
\def\jnumber#1#2{\thispagestyle{empty} \noi\unitlength=1mm
        \begin{picture}(178,10)
            \put(177,15){\llap{\large\it Grav. Cosmol. No.\,#1, #2}}
                    \end{picture}}
\newcommand{\Title}[1]{\noi {{\Large\bf #1}}\\[1ex]}
\newcommand{\Author}[2]{\noi{\bf #1}\\[2ex]\noi{\normalsize\it #2}\\}

\newcommand{\Abstract}[1]{\vskip 2mm \begin{center}
        \parbox{16.4cm}{\small\noi #1} \end{center}\medskip}

\newcommand{\foom}[1]{\protect\footnotemark[#1]}
\newcommand{\foox}[2]{\footnotetext[#1]{#2}\addtocounter{footnote}{1}}

\def\Talk{\foox 1 {Talk given at the International Conference RUDN-10,
       June 28 --- July 3, 2010, PFUR, Moscow}}

\def\nq{\hspace*{-1em}}
\def\nqq{\hspace*{-2em}}

\def\cm{\hspace*{1cm}}
\def\inch{\hspace*{1in}}

\def\Jl#1#2{#1 {\bf #2},\ }

\def\ApJ#1 {\Jl{Astroph. J.}{#1}}
\def\CQG#1 {\Jl{Class. Quantum Grav.}{#1}}
\def\DAN#1 {\Jl{Dokl. AN SSSR}{#1}}
\def\GC#1 {\Jl{Grav. Cosmol.}{#1}}
\def\GRG#1 {\Jl{Gen. Rel. Grav.}{#1}}
\def\JETF#1 {\Jl{Zh. Eksp. Teor. Fiz.}{#1}}
\def\JETP#1 {\Jl{Sov. Phys. JETP}{#1}}
\def\JHEP#1 {\Jl{JHEP}{#1}}
\def\JMP#1 {\Jl{J. Math. Phys.}{#1}}
\def\NPB#1 {\Jl{Nucl. Phys. B}{#1}}
\def\NP#1 {\Jl{Nucl. Phys.}{#1}}
\def\PLA#1 {\Jl{Phys. Lett. A}{#1}}
\def\PLB#1 {\Jl{Phys. Lett. B}{#1}}
\def\PRD#1 {\Jl{Phys. Rev. D}{#1}}
\def\PRL#1 {\Jl{Phys. Rev. Lett.}{#1}}

\def\lal{&&\nqq {}}

\def\eqs{Eqs.\,}
\def\beq{\begin{equation}}
\def\eeq{\end{equation}}
\def\bear{\begin{eqnarray}}
\def\bearr{\begin{eqnarray} \lal}
\def\ear{\end{eqnarray}}
\def\earn{\nonumber \end{eqnarray}}

\def\nnn{\nonumber\\ \lal }

\def\yy{\\[5pt] {}}
\def\yyy{\\[5pt] \lal }

\def\d{\partial}

\def\const{{\rm const}}
\def\eps{\varepsilon}

\begin{document}
\twocolumn[
\jnumber{1}{2011}

\Title{Exact solution of the relativistic magnetohydrodynamic equations \yy
       in the background of a plane gravitational wave\yy
       with combined polarization\foom 1}

\Author{A. A. Agathonov and Yu. G. Ignatyev}
    {Kazan State Pedagogical University, Mezhlauk str. 1,
    Kazan 420021, Russia}

\Abstract
 {We obtain an exact solution of the self-consistent relativistic
  magnetohydrodynamic equations for an anisotropic magnetoactive
  plasma in the background of a plane gravitational wave metric (PGW)
  with an arbitrary polarization. It is shown that, in the linear
  approximation in the gravitational wave amplitude, only the $\mathbf{e_+}$
  polarization of the PGW interacts with a magnetoactive plasma.}

] 
\Talk

\section {Introduction}

  In a series of previous articles by one of the authors (see, e.g., [1--3]
  a theory of {\it gravimagnetic shock
  waves} in a homogeneous magnetoactive plasma has been developed. The
  essence of this phenomenon is that a magnetized plasma in anomalously
  strong magnetic fields drifts under the action of gravitational waves
  (GWs) in the GW propagation direction under the condition that the
  wave amplitude is large enough, and, on a certain wave front, the plasma
  velocity tends to the speed of light. Its energy density and the intensity
  of the frozen-in magnetic field then tend to infinity. In the subsequent
  papers this effect was proved on the basis of the kinetic theory, and
  the possibility of using this mechanism as an effective tool for detecting
  GWs from astrophysical sources was also shown. However, in all cited
  papers, a monopolarized gravitational wave was considered. In the present
  paper we consider the action of a GW with combined polarization on a
  magnetoactive plasma.

\section{Self-consistent RMHD equations in a gravitational field}

  In [1], under the assumption that the dynamic velocity of the
  plasma ($v^i$) is equal to that of the electromagnetic field\footnote
    {The index ``$p$'' refers to the plasma, the index ``$f$'' to
    the field, the comma denote a covariant derivative. The dynamic
        velocity of any kind of matter is, by definition, a timelike unit
    eigenvector of the energy-momentum tensor of this matter
    \cite{ig-Synge}.}
\beq      \label{ig-eq_vel}
    \stackrel{p}{T}_{ij}v^j=\eps_p v_i;\quad \stackrel{f}{T}
        _{ij}v^j=\eps_f v_i, \quad (v,v)=1,
\eeq
  a full self-consistent set of relativistic
  magneto\-hy\-d\-ro\-dy\-na\-mic equations for a magnetized plasma in
  arbitrary gravitational field has been obtained. It consists of the
  Maxwell equations of the first group
\beq      \label{ig-1Maxwell}
    \stackrel{*}{F}\ \!\!\!^{ik}_{~~,k}=0
\eeq
  with the necessary and sufficient condition
\bear      \label{ig-I_inv}
    {\rm Inv}_1&=&F_{ij}F^{ij}=2H^2>0,
\\                \label{ig-II_inv}
    {\rm Inv}_2&=&\stackrel{*}{F}_{ij}F^{ij}=0,
\ear
  the Maxwell equations of the second group\footnote
    {$c = G = \hbar = 1$}:
\beq            \label{ig-2Maxwell}
    F^{ik}_{~~,k} = -4\pi J^i_{\rm{dr}}
\eeq
  with a spacelike {\it drift current}
\beq      \label{ig-Jdr}
    J^i_{\rm{dr}}=-\frac{2F^{ik} \stackrel{p}{T}\
    \!\!\!^{l}_{k,l}}{F_{jm}F^{jm}},\quad (J_{\rm{dr}},J_{\rm{dr}}) < 0
\eeq
  and a conservation law for the total energy-momentum of the system
\beq      \label{ig-Tik,k}
    T^{ik}_{~,k}=\stackrel{p}{T}\ \!\!\!^{ik}_{~,k}+ \stackrel{f}{T}\
        \!\!\!^{ik}_{~,k}=0.
\eeq

  The energy-momentum tensor (EMT) of the elec\-t\-ro\-mag\-ne\-tic field,
  in the case of a coincidence of the plasma's and the field's
  dynamic velocities (\ref{ig-eq_vel}), is expressed through a pair
  of vectors, $v$ and $H$ \cite{ig-Ign95}:
\beq      \label{ig-T_f_H}
    \stackrel{f}{T}\ \!\!\!^{i}_k = -\frac{1}{8\pi}\left[(\delta^i_k-2v^i
        v_k)H^2+2H^i H_k)\right].
\eeq
  The EMT of a relativistic anisotropic mag\-ne\-to\-ac\-tive plasma in
  gravitational and magnetic fields is (see, e.g., \cite{ig-IgnGor97})
\beq      \label{ig-T_p}
    \stackrel{p}{T}\ \!\!\!^{ij}=(\eps +p_\perp)v^iv^j-p_\perp
        g^{ij}+(p_\parallel -p_\perp)h^ih^j,
\eeq
  where $h^i = H^i/H$ is the spacelike unit vector of the magnetic field
  ($(h,h)=-1$); $p_\perp$ and $p_\parallel$ are the plasma pressures in the
  directions orthogonal and parallel to the magnetic field, respectively.

\section{Solving the RMHD equations in the PGW metric}

  Consider a solution of the Cauchy problem of the self-consistent RMHD
  equations in the background of a vacuum gravitational-wave metric (see,
  e.g., \cite{ig-torn})\footnote
    {$\beta(u)$ and $\gamma(u)$ are the amplitudes of the polarizations
     $\mathbf{e}_+$ and $\mathbf{e}_\times$, respectively; $u
     =(t-x^1)/\sqrt{2}$ is the retarded time, $v=(t + x^1)/\sqrt{2}$ is
     the advanced time. The PGW amplitudes are arbitrary functions of
     the retarded time $u$, and $L(u)$ is a background factor of the PGW.}:
\bearr
    d s^{2} = 2 du dv  - L^{2} \biggl[\cosh 2\gamma\biggl(e^{2\beta}
        (dx^{2})^{2}
\nnn \cm      \label{ig-01}
     + 2 e^{-2\beta}( dx^{3})^{2}\biggr) -\sinh 2\gamma dx^2 dx^3\biggr],
\ear
  with homogeneous initial conditions on the null hypersurface $u=0$:
\bearr        \label{ig-03}
    \beta(u \leq 0)=0; \quad \beta'(u\leq 0)=0;\quad L(u \leq 0)=1,
\nnn
\ear             
  We assume the following:
\begin{itemize}
\item
     the plasma is homogeneous and at rest:
\bearr
    v^v(u\leq 0)= v^u(u\leq 0) = 1/\sqrt{2};
\nnn
    v^{2} =v^{3}=0; \qquad \eps(u \leq 0)=\stackrel{0}{\eps};
\nnn          \label{ig-04a}
    p_\parallel(u \leq 0) = \stackrel{0}{p}_\parallel; \qquad
    p_\perp(u \leq 0) = \stackrel{0}{p}_\perp;
\ear
\item
    a homogeneous magnetic field is directed in the $(x^1,x^2)$ plane:
\bearr
    H_1(u \leq 0)=\stackrel{0}{H} \cos\Omega\,;
\nnn
    H_2(u \leq 0)=\stackrel{0}{H} \sin\Omega\,;
\nnn       \label{ig-05}
    H_3(u \leq 0) = 0, \qquad  E_i(u \leq 0) = 0,
\ear
  where $\Omega$ is the angle between the axis $0x^{1}$ (the PGW propagation
  direction) and the magnetic field ${\bf H}$.
\end{itemize}

  The metric (\ref{ig-01}) admits the group of isometries $G_{5}$,
  associated with three linearly independent (at a point) Killing vectors
\beq       \label{ig-02}
    \mathop{\xi^{i}}\limits_{(1)} =\delta^{i}_{v}\,; \qquad
    \mathop{\xi^{i}}\limits_{(2)} = \delta^{i}_{2}\,; \qquad
    \mathop{\xi^{i}}\limits_{(3)} = \delta^{i}_{3}\,.
\eeq
  Due to their existence in the metric (\ref{ig-01}), all geometric objects,
  including the Christoffel symbols, the Riemann tensor, the Ricci tensor
  and consequently the EMT of a magnetoactive plasma, are automatically
  conserved at motions along the Killing directions:
\beq      \label{ig-symmetric}
    \mathop{\mathrm{L}}\limits_{\xi_\alpha}g_{ij}=0\ \Rightarrow\
    \mathop{\mathrm{L}}\limits_{\xi_\alpha}R_{ij}=0\ \Rightarrow\
    \mathop{\mathrm{L}}\limits_{\xi_\alpha}T_{ij}=0,
\eeq
  where $\mathop{\mathrm{L}}\limits_{\xi}T_{ij}$ is a Lie derivative in the
  direction of $\xi$. We further require that the EMTs of the plasma
  $\stackrel{p}{T}_{ij}$ and the electromagnetic field $\stackrel{f}{T}_{ij}$
  inherit the symmetry separately. Thus all observed physical quantities
  $\mathbf{P}$ {\em inherit the symmetry of the metric} (\ref{ig-01}):
\beq                                 \label{ig-07}
    \mathop{\mathrm{L}}\limits_{\xi_\alpha} {\bf P} =0 \cm
    (\alpha =\overline{1,3}),
\eeq
  i.e., taking into account the explicit form of the Killing vectors
  (\ref{ig-02}),
\bearr            \label{ig-08}
    p=p(u), \quad \eps=\eps(u), \quad v^{i}=v^{i}(u);
\yyy               \label{ig-09}
    F_{ik}=F_{ik}(u), \quad H_i=H_i(u), \quad h_i=h_i(u).
\ear

  The vector potential agreeing with the initial conditions (\ref{ig-05}) is
\bearr
    A_{v} = A_{u} = A_{2} = 0;
\nnn                          \label{ig-06}
    A_3 = \stackrel{0}{H} (x^1 \sin\Omega - x^2\cos\Omega); \qquad (u\leq 0).
\ear
  In the presence of a PGW, the vector potential becomes
\bearr
    A_2=A_v=A_u=0;
\nnn          \label{ig-Ai}
    A_3=\stackrel{0}{H}\left(\frac{1}{\sqrt{2}}(v-\psi(u))
        \sin\Omega-x^2\cos\Omega \right),
\ear
  where $\psi(u)$ is an arbitrary function of the retarded time,
  satisfying the initial condition
\beq      \label{ig-phi0}
    \psi (u\leq 0) = u.
\eeq
  Thus the magnetic field freezing-in condition in the plasma reduces to
  the two equalities
\bearr      \label{ig-vi}
    v^3 = 0,
\nnn
    \frac{1}{\sqrt{2}}(v_v\psi'-v_u)\sin\Omega+v^2\cos\Omega=0.
\ear
  The covariant components of the vector of magnetic field intensity
  relative to the Maxwell tensor are
\bearr      \label{ig-Hv}
    H_v=-\frac{\stackrel{0}{H}}{L^2} \left(v_v
    \cos\Omega+\frac{1}{\sqrt{2}} v^2 \sin\Omega \right)
\\ \lal            \label{ig-Hu}
    H_u = \frac{\stackrel{0}{H}}{L^2}
      \left( v_u \cos\Omega- \frac{1}{\sqrt{2}}v^2 \psi' \sin\Omega \right),
\\ \lal              \label{ig-H2}
    H_2 = -\frac{1}{\sqrt{2}} \stackrel{0}{H} \cosh2\gamma e^{2\beta}
        \sin\Omega (v_v \psi'+ v_u ),
\\  \lal                   \label{ig-H3}
    H_3 = \frac{1}{\sqrt{2}} \stackrel{0}{H} \sinh2\gamma
        \sin\Omega (v_v \psi'+ v_u ).
\ear
  The magnetic field intensity squared is
\beq       \label{ig-33}
    H^2 = \frac{\stackrel{0}{H}\ \!\!\!^{2}}{L^4} (L^2 \psi'
    \cosh2\gamma e^{2\beta} \sin^2\Omega + \cos^2\Omega )\,.
\eeq

  Using (\ref{ig-Hv})-(\ref{ig-33}), the normalization relation for the
  velocity vector can be written in the equivalent form
\bearr        \label{ig-34}
    \left[ v_v \cos\Omega + v_2 \frac{1}{\sqrt{2}} \sin\Omega \right]^2
\nnn \cm
    = \frac{H^2}{\stackrel{0}{H}\ \!\!\!^{2}}
    v^2_v L^4 - \frac{\sin^2 \Omega}{2} L^2 \cosh2\gamma e^{2\beta}\,.
\ear
  The components of the drift current are
\beq      \label{ig-curr}
    J^i_{\rm{dr}} = -\frac{1}{4\pi L^2}\d_u (L^2 F^{iu}).
\eeq
  Then,
\bearr          \label{ig-J^v}
    J^v_{\rm{dr}} = J^u_{\rm{dr}} = 0,
\yyy        \label{ig-J^2}
    J^2_{\rm{dr}} = -\frac{\stackrel{0}{H}\sin\Omega}{2\sqrt{2}\pi L^2}
        \cosh2\gamma\cdot\gamma' ,
\yyy      \nq        \label{ig-J^3}
    J^3_{\rm{dr}} = -\frac{\stackrel{0}{H}\sin\Omega e^{2\beta}}
    {2\sqrt{2}\pi L^2}(\sinh2\gamma\cdot\gamma'+\cosh2\gamma\cdot\beta').
\ear
  Because of existence of the isometries (\ref{ig-02}), we obtain the
  following integrals \cite{ig-Ign95}:
\beq        \label{ig-35}
    L^2 \mathop{\xi}\limits_{(\alpha)}{}^i T_{v i} = C_a = \const
    \qquad (\alpha = \overline{1,3})\,.
\eeq

  We consider only the case of {\it transverse PGW propagation\/}
  ($\Omega=\pi/2$). Then, substituting the expres\-si\-ons for the
  plasma and electromagnetic field EMT into the integrals (\ref{ig-35}),
  using the relations (\ref{ig-H3})-(\ref{ig-34}) and also the initial
  conditions (\ref{ig-03}), we bring the integrals of motion to the form
\bearr              \label{ig-C1}
    2 L^2 (\eps + p_\parallel) v_v^2 - (p_\parallel - p_\perp)
    \frac{\stackrel{0}{H}\ \!\!\!^{2}}{H^2} \cosh2\gamma e^{2\beta}
\nnn \inch
    = (\stackrel{0}{\eps} + \stackrel{0}{p}) \Delta(u),  
\yyy            \label{ig-C2}
    L^2 (\eps + p_\parallel) v_v v_2 = 0,
\yyy            \label{ig-C3}
    L^2 (\eps + p_\parallel) v_v v_3 = 0,
\ear
  where
\beq        \label{ig-38}
    \stackrel{0}{p} = \stackrel{0}{p}_\perp,
\eeq
  and the so-called {\it governing function of the GMSW} is introduced:
\beq        \label{ig-40}
    \Delta(u) =   1 - \alpha^2 (\cosh2\gamma e^{2\beta} - 1)\,,
\eeq
  with the {\it dimensionless parameter} $\alpha^2$,
\beq        \label{ig-alpha}
    \alpha^2 = \frac{\stackrel{0}{H}\ \!\!\!^{2}}{4\pi
            (\stackrel{0}{\eps} + \stackrel{0}{p})}\,.
\eeq

  Solving (\ref{ig-C1}) with respect to $v_v$, we obtain expressions for the
  components of the velocity vector as functions of the scalars $\eps$,
  $p_\parallel$, $p_\perp$, $\psi'$ and explicit functions of the retarded
  time:
\bearr        \label{ig-Vv}
    v_v^2 =\frac{(\stackrel{0}{\eps}\stackrel{0}{p})}
        {2L^2(\eps + p_\parallel)} \Delta(u)      \inch
\nnn \cm
    + \frac{(p_\parallel - p_\perp)}{(\eps + p_\parallel)}
    \frac{\stackrel{0}{H}\ \!\!\!^{2}}{H^2} \frac{\cosh2\gamma
        e^{2\beta}}{2L^2}.
\ear
  From (\ref{ig-C2}), (\ref{ig-C3}) we get:
\beq        \label{ig-V2}
    v_2 = v_3=0 \,.
\eeq
  We obtain the component $v_u$ from the normalization relation for the
  velocity vector, using (\ref{ig-Vv}) and (\ref{ig-V2}):
\beq        \label{ig-Vu}
    v_u = \frac{1}{2 v_v}\,,
\eeq
  and from the freezing-in condition (\ref{ig-vi}) we get the value of the
  derivative of potential $\psi'$:
\beq        \label{ig-psi}
    \psi' = \frac{1}{2 v_v^2}.
\eeq
  Using it, the scalar $H^2$ is determined from the relation (\ref{ig-33}):
\beq        \label{ig-H^2(perp)}
    H^2 = \frac{\stackrel{0}{H}\ \!\!\!^{2}}{L^2} \frac{\cosh2\gamma
        e^{2\beta}}{2 v_v^2}.
\eeq
  From the RMHD set of equations it is possible to obtain the following
  differential equation in the PGW metric:
\bearr          \label{ig-47}
    L^2 \eps' v_v + (\eps + p_\parallel)(L^2 v_v)'   \cm\cm
\nnn \cm
    + \frac{1}{2}L^2 (p_\parallel - p_\perp) v_v (\ln H^2)' = 0\,.
\ear
  To solve this equation, it is necessary to impose two additional relations
  between the functions $\eps$, $p_\parallel$, and $p_\perp$, i.e., an
  equation of state:
\beq        \label{ig-48}
    p_\parallel = f(\eps)\,; \quad p_\perp = g(\eps)\,.
\eeq

\section{Barotropic equation of state}

  Consider a barotropic equation of state of the anisotropic plasma, where
  the relations (\ref{ig-48}) are linear:
\beq       \label{ig-49}
    p_\parallel = k_\parallel \eps \,; \quad p_\perp = k_\perp \eps\,.
\eeq
  Equation (\ref{ig-47}) is easily integrated under the conditions
 (\ref{ig-49}), and we get one more integral:
\beq        \label{ig-50}
    \eps (\sqrt{2} L^2 v_v)^{(1 + k_\parallel)} H^{(k_\parallel -
    k_\perp)} = \stackrel{0}{\eps} \stackrel{0}{H}\
        \!\!\!^{(k_\parallel - k_\perp)}\,.
\eeq
  In the case of a barotropic equation of state under the conditions
  (\ref{ig-49}), substitution of (\ref{ig-H^2(perp)}) into (\ref{ig-Vv})
  results in
\beq        \label{ig-54}
    v^2_v = \frac{1}{2}\frac{\stackrel{0}{\eps}}{L^2 \eps}\Delta (u)\,.
\eeq

  Substituting (\ref{ig-H^2(perp)}) and (\ref{ig-54}) into (\ref{ig-50}), we
  obtain a closed equation with respect to the variable $\eps$,
  whose solution gives:
\bearr        \label{ig-bar_E}
    \eps = \stackrel{0}{\eps} \Big[ \Delta^{1+k_\perp}
        L^{2(1+k_\parallel)} (\cosh2\gamma
        e^{2\beta})^{k_\parallel-k_\perp} \Big]^{-g_\perp},
\yyy            \label{ig-bar_Vv}
    v_v = \frac{1}{\sqrt{2}} \left[ \Delta L^{(k_\parallel+k_\perp)}
    (\cosh 2\gamma e^{2\beta})^{\frac{k_\parallel-k_\perp}{2}}
    \right]^{g_\perp},
\yyy            \label{ig-bar_H} \displaystyle
    H = \stackrel{0}{H} \left[ \Delta L^{(1+k_\parallel)}
    (\cosh 2\gamma e^{2\beta})^{-\frac{1-k_\parallel}{2}}
        \right]^{-g_\perp}\,,
\ear
  where
\beq        \label{ig-58}
    g_\perp = \frac{1}{1 - k_\perp} \in [1, 2]\,.
\eeq
  In particular, for an ultrarelativistic plasma with zero parallel pressure,
\beq        \label{ig-59}
    k_\parallel \to 0\,; \quad k_\perp \to \frac{1}{2}
\eeq
  we obtain from (\ref{ig-bar_E})--(\ref{ig-58}):
\bearr        \label{ig-60}
    v_v = \frac{1}{\sqrt{2}} L \Delta^2 (\cosh2\gamma e^{2\beta})^{-1/2},
\\ \lal         \label{ig-61}
    \eps = \stackrel{0}{\eps} L^{-4} \Delta^{-3}
        (\cosh2\gamma e^{2\beta}),
\yyy            \label{ig-62}
    H = \stackrel{0}{H} L^{-2} \Delta^{-2} (\cosh2\gamma e^{2\beta})\,.
\ear

\section{The energy balance equation}

  In \cite{ig-Ign95}, it has been shown that the singular state, which
  exists in a magnetized plasma under the condition $2 \beta_0 \alpha^2 > 1$
  on the hypersurface
\beq        \label{ig-62}
    \Delta(u_*) = 0\,,
\eeq
  is removed using the back reaction of the magnetoactive plasma on the GW.
  That leads to efficient absorption of GW energy by the plasma and a
  restric\-ti\-on on the GW amplitude. A qualitative analysis of this
  situation can be carried out using a simple model of energy balance
  proposed in \cite{ig-Ign96}. The energy flow of the magnetoactive plasma
  is directed along the PGW propagation direction, i.e., along the $x^1$
  axis.  Let $\beta_*(u)$ and $\gamma_*(u)$ be the vacuum PGW amplitudes.
  In the WKB approxi\-ma\-ti\-on,
\beq      \label{ig-WKB}
    8\pi\eps \ll \omega^2\,,
\eeq
  where $\omega$ is the characteristic PGW frequency and $\eps$ is the
  matter energy density, all functions still depend on the retarded time
  only (see \cite{ig-IgnBal81}). Thus $\beta(u)$ and $\gamma(u)$ are the
  PGW amplitudes subject to absorption in plasmas. The local energy
  conservation law should be satisfied:
\beq      \label{ig-eq_T_41}
    T^{41}(\beta,\gamma) + \stackrel{g}{T}{}^{41}(\beta,\gamma) =
        \stackrel{g}{T}{}^{41}(\beta_*,\gamma_*)\, ,
\eeq
  where $\stackrel{g}{T}{}^{41}(\beta,\gamma)$ is the energy flow of a
  weak GW in the direction $0x^{1}$ (see \cite{ig-land}).

  In the case of transversal PGW propagation and with a barotropic equation
  of state of an anisotropic plasma, using the solutions of
  magnetohy\-d\-ro\-dy\-na\-mics and \eqs (\ref{ig-bar_E}),
  (\ref{ig-bar_Vv}), (\ref{ig-bar_H}) with the dimensionless parameter
  $\alpha^2$ (\ref{ig-alpha}), one can obtain the energy balance equation in
  the form
\bearr               \label{ig-eq_T_temp2}
    \frac{\stackrel{0}{H}\ \!\!\!^{2}}{4}\left( \Delta^{-4 g_\perp} -
    1\right) \left(\frac{1}{\alpha^2} + 1\right)
\nnn \cm
    + (\gamma')^2 + (\beta')^2 = (\gamma'_*)^2 + (\beta'_*)^2.
\ear
  Since, in a linear approximation by smallness of the amplitudes $\beta$
  and $\gamma$, the governing function (\ref{ig-40}) does not depend on the
  function $\gamma(u)$,
\beq        \label{ig-65}
    \Delta(u) = 1 - 2 \alpha^2 \beta  +O(\beta^2,\gamma^2)\,,
\eeq
  and the functions $\beta(u)$, $\gamma(u)$ are arbitrary and functionally
  indepen\-dent, then, up to $\beta^2, \gamma^2$, the relation
  (\ref{ig-eq_T_temp2}) can be split into two indepen\-dent parts:
\bearr        \label{ig-b}
    2\stackrel{0}{H}\ \!\!\!^{2} g_\perp (1+\alpha^2)\beta+(\beta')^2
            =(\beta'_*)^2,
\yyy
    (\gamma')^2=(\gamma'_*) ^2.      \label{ig-g}
\ear
  Here, according to the meaning of the local energy balance equation, we
  consider short gravitational waves (\ref{ig-WKB}), so we can neglect
  the squares of the PGW amplitudes as compared with the squares of their
  derivatives with respect to the retarded time. Thus, according to
  (\ref{ig-g}),
\beq        \label{ig-66}
    \gamma_*(u) = \gamma(u),
\eeq
  i.e., in the linear approximation, a weak gravitational wave with the
  polarization ${\bf e}_{\times}$ does not interact with a magnetized plasma.
  This coincides with the conclu\-si\-on of the paper \cite{ig-IgnKhu86}.
  Thus the energy balance equation takes the form obtained in
  \cite{ig-IgnGor97}:
\beq        \label{ig-77}
    \dot{\Delta}^2 + \xi^2 \Upsilon^2 \Bigl[\Delta^{- 4g_\perp} - 1
        \Bigr] = \Upsilon^2 \sin^2(s),
\eeq
  where $\xi^2$ is the so-called {\it first parameter of the GMSW}
  \cite{ig-Ign96}:
\bearr        \label{ig-71}
    \xi^2 = \frac{\stackrel{0}{H}\ \!\!\!^{2}}{4 \beta^2_0 \omega^2},
\yyy            \label{ig-72}
    \Upsilon = 2\alpha^2\beta_0
\ear
  --- {\it the second GMSW parameter}. The dot denotes differentiation
  with respect to the dimensionless time variable $s$,
\beq       \label{ig-69}
    s = \sqrt{2} \omega u.
\eeq

\section{Conclusion}

  Thus we have obtained a generalization of the results of
  \cite{ig-Ign95}-\cite{ig-IgnGor97} to gravitational waves
  with two polarizations and showed that, in the linear approximation,
  the polarization $\mathbf{e}_\times$ does not interact with a
  magnetized plasma. This justifies the applicability
  of the previously obtained results for arbitrarily
  polarized gravitational waves.

\end{document}